\documentclass[manuscript,screen]{acmart}
\AtBeginDocument{%
  }

\setcopyright{acmlicensed}
\copyrightyear{2025}
\acmYear{2025}
\acmDOI{XXXXXXX.XXXXXXX}
\acmISBN{978-1-4503-XXXX-X/2018/06}

\usepackage{amsmath,amsfonts}
\usepackage{algorithmic}
\usepackage{graphicx}
\usepackage{textcomp}
\usepackage{url}
\usepackage{balance}
\usepackage{xcolor}
\usepackage{multirow}
\usepackage{multicol}
\usepackage{listings}
\newcommand{\rev}[1]{\textcolor{black}{#1}}

\begin{document}

\title{ReCatcher: Towards LLMs Regression Testing for Code Generation}

\author{Altaf Allah Abbassi}
\email{altaf-allah.abbassi@polymtl.ca}
\affiliation{%
  \institution{Polytechnique Montreal}
  \city{Montreal}
  \state{Quebec}
  \country{Canada}
}
\author{Leuson Da Silva}
\email{leuson-mario-pedro.da-silva@polymtl.ca}
\affiliation{%
  \institution{Polytechnique Montreal}
  \city{Montreal}
  \state{Quebec}
  \country{Canada}
}
\orcid{0000-0002-9086-9038}

\author{Amin Nikanjam}
\affiliation{%
  \institution{Huawei Distributed Scheduling and Data Engine Lab}
  \city{Montreal}
  \country{Canada}
}
\email{amin.nikanjam@h-partners.com}
\orcid{0000-0002-0440-6839}
\authornote{Work done while at Polytechnique Montreal.}

\author{Foutse Khomh}
\email{foutse.khomh@polymtl.ca}
\affiliation{%
  \institution{Polytechnique Montreal}
  \city{Montreal}
  \state{Quebec}
  \country{Canada}
}

\renewcommand{\shortauthors}{Abbassi et al.}

\begin{abstract}
Large Language Models (LLMs) for code generation evolve rapidly through fine-tuning, merging, or new model releases. 
However, such updates can introduce regressions—not only in correctness but also in code quality and performance. 
To address this, we present \textit{ReCatcher}, a regression testing framework for Python code generation. 
\textit{ReCatcher} systematically compares two LLMs—typically a current model and a candidate update—across three dimensions: logical correctness, static code quality, and execution performance. 
We apply \textit{ReCatcher} to assess regressions across three update scenarios—fine-tuning, merging, and model release—using CodeLlama, DeepSeek-Coder, and GPT-4o. Our evaluation shows that fine-tuning with cross-language datasets increases syntax errors by up to 12\%. 
Merging with general-purpose models like Llama2 leads to regressions in correctness by up to 18\%. 
GPT-4o introduces regressions of up to 50\% in handling missing imports compared to GPT-3.5-turbo, while GPT-4o-mini suffers up to 80\% performance degradation in execution time versus GPT-4o. 
Overall, logical correctness, performance, and error handling (e.g., syntax errors and missing imports) are the most regression-prone areas. Comparing \textit{ReCatcher} with baseline solutions, it presents better and consistent accuracy across logical and performance aspects. 
\textit{ReCatcher} highlights the importance of systematic regression evaluation before adopting new models, while assisting researchers and practitioners in making more informed update decisions.
\end{abstract}




\maketitle

\section{Introduction}
LLMs have demonstrated code generation capability across multiple programming languages, including Python, Java, and C/C++~\cite{liu2024your}. 
With their increasing use in software development~\cite{jahic2024state}, there has been a shift toward models specifically trained for code generation~\cite{jiang2024survey}, ranging from proprietary models such as GPT-4 to open-source models such as CodeLlama~\cite{touvron2023llama} and DeepSeek-Coder~\cite{guo2024deepseek}. 
To further improve code generation capability, new LLMs are frequently introduced either by releasing new model versions (such as GPT-4 succeeding GPT-3.5) or by adapting existing models with techniques such as merging or fine-tuning.
Fine-tuning enables models to adapt to specific programming language or tasks by further training on domain-specific data~\cite{li2024fine}. 
For example, Li et al.~\cite{li2024fine} fine-tuned GPT-J to enhance their proficiency in  C/C++. 
On the other hand, merging combines multiple models' strengths, as demonstrated by Liu et al.~\cite{liu2025sensmergingsensitivityguidedparameterbalancing}, who integrated models with mathematical reasoning and general logic capabilities to improve code generation.

Regarding assessing LLM-generated code, the evaluation primarily focuses on functional correctness, often measured by metrics like pass@k~\cite{jiang2024survey}.
However, focusing solely on correctness fails to capture critical code aspects, particularly when integrating LLM-generated code into real-world software development~\cite{zheng2024beyond}.
Studies have shown that LLM-generated code is prone to various code issues, such as maintainability, performance, readability, and security~\cite{zheng2024beyond, MORADIDAKHEL2023111734, fu2023security, 10.1145/3643991.3645071, ouedraogo2024test}.
These issues contribute to inefficiencies, which we define as any factor that negatively impacts code quality. Inefficient code contains one or more inefficiencies, while efficient code remains free of them.
 
Inefficiencies can degrade software quality~\cite{jiang2024survey}. As LLMs for code generation evolve, ensuring that new versions do not introduce or amplify inefficiencies is essential. This need establishes regression testing as a critical practice for LLM evaluation~\cite{ma2024my}. 
In traditional Software Engineering (SE), regression testing is typically performed by re-executing unit tests to verify that changes do not introduce unintended failures~\cite{biswas2011regression}. However, in the case of LLM-assisted programming, this approach is challenging due to the nondeterministic behavior of LLMs that prevents the application of unit tests. 
The same prompt can generate different outputs, making it difficult to define expected results and reliably detect regressions~\cite{ouyang2025empirical}.

While efforts in LLMs regression testing are limited, RETAIN~\cite{dixit2024retain} was designed as a general-purpose LLM regression testing framework, focusing primarily on textual differences between LLMs outputs. 
However, textual differences alone cannot effectively assess code generation because code is not just text; it has structural, semantic, and executional properties that impact both functional and non-functional aspects~\cite{naik2024limitations}. 
This limitation makes RETAIN an unsuitable regression testing framework for code generation. 
\rev{Aiming to address this gap, Zheng et al. \cite{zheng2024beyond} propose and evaluate RACE, a framework to assess LLM-generated code. Although RACE considers multiple dimensions—such as correctness, readability, maintainability, and efficiency—it primarily targets general aspects of code quality, relying on conventional standard metrics, rather than challenges specific to LLM-generated code \cite{abbassi2025unveiling}. 
Furthermore, RACE lacks structured test formulations, impacting its ability to systematically assess and reproduce results across different inefficiency subcategories.}

To address this gap, we propose \textit{ReCatcher}, to the best of our knowledge, the first LLM regression testing framework for code generation. 
\textit{ReCatcher} systematically compares the code generation capability of two LLMs—the current model in use and a potential model for update—and generates a comprehensive regression report.
Our framework tests regression across three key aspects: logical correctness, performance, and static code issues (including readability, maintainability, and errors), leveraging a taxonomy of inefficiencies in LLM-generated code~\cite{abbassi2025unveiling}. 
\textit{ReCatcher} integrates widely validated software testing tools: unit tests for correctness \cite{fraser2011evosuite}, static analysis for static code issues \cite{hassan2024evaluating, pylint}, and profiling tools for performance \cite{traini2021software}. 
By systematically applying these established methods and comparing LLM-generated code snippets across models, ReCatcher provides a robust and reproducible assessment of regressions in LLM-generated code.
\textit{ReCatcher} supports specifically Python, considering its popularity and wide adoption~\cite{spectrum2024}. 

We used \textit{ReCatcher} to assess the impact of three common model update scenarios on LLMs' code generation capability: fine-tuning~\cite{tsai2024code}, merging~\cite{dehghan2024mergerepair}, and model release, which involves introducing a new version within a model family (e.g., the release of GPT-4o-mini~\cite{openai2024gpt4omini}).
\rev{While \textit{merging} and \textit{fine-tuning} are widely adopted update strategies in software engineering tasks—particularly for code generation~\cite{jiang2024survey, zheng2023survey}—LLMs and their evolving versions are increasingly used as code assistants~\cite{tian2023chatgpt, da2025llms}, highlighting the need for a deeper investigation into their impact on code inefficiencies and overall quality.}
Our assessment covers both open-source and proprietary LLMs. 
For open-source models, we tested regression when updating to the fine-tuned and merged variants of the widely used CodeLlama \cite{roziere2023code} and DeepSeek-Coder \cite{guo2024deepseek}. The selected variants were retrieved from Hugging Face\cite{HuggingFaceHub}, prioritizing those with the highest number of downloads.  
For proprietary models, we assessed regression when updating from GPT-3.5-turbo to GPT-4o and from GPT-4o to GPT-4o-mini. The GPT family is among the most widely used model families for code generation \cite{jiang2024survey}.
To ensure a comprehensive evaluation, we used two benchmarks: HumanEval+~\cite{liu2024your} (which consists of algorithmic problem-solving tasks) and BigCodeBench~\cite{zhuo2024bigcodebench} (which covers real-world software development tasks). 

Across all tested scenarios, the most prevalent regressions occurred in logical correctness, errors (syntax error and missing declaration/import), and performance, while readability and maintainability were relatively stable.
Based on our results, we recommend prioritizing testing these prevalent aspects before model adoption. 
Regarding the different adaptation techniques, we observe that fine-tuning models on datasets from different programming languages led to syntax errors, resulting in a regression of up to 12\%. We observe that LLMs merging can either mitigate or amplify regressions depending on the merged model specialization and training objective. For instance, merging with a general-purpose LLM (a non-coding optimized LLM such as Llama2 \cite{touvron2023llama}) introduced regression across various code quality aspects, such as logical correctness, by up to 18\%. 
Finally, regarding model releases, GPT-4o exhibited regression in handling missing declaration/import compared to GPT-3.5-turbo up to 50\%. 
Additionally, GPT-4o-mini showed performance regressions compared to GPT-4o; specifically, it regressed by 80\% for execution time.
For merging, selecting models with coding capabilities helps mitigate regressions. Fine-tuning on datasets from different programming languages increases syntax errors. 

\rev{We also compare the accuracy of ReCatcher against baseline solutions. 
Specifically, we looked for pairs of models with minimal variation in logical correctness and performance aspects, as established in our prior study, selecting CodeLlama and its finetuned version on HumanEval+. 
Rather than repeating the full evaluation, by targeting such a challenging scenario, we aim to reveal subtle quality differences and stress-test ReCatcher in challenging scenarios where baseline methods often struggle to capture nuanced variations.
Using LLMs as baseline judges, alongside standard code quality metrics, we observed that ReCatcher achieves superior and more consistent accuracy across both logical and performance dimensions. 
Moreover, it effectively addresses static code issues commonly associated with LLM-generated outputs, highlighting its robustness in identifying inefficiencies specific to LLM code-generated.} 

\rev{By providing a structured and comprehensive regression report, ReCatcher enables developers and researchers to assess trade-offs introduced by model updates and make informed decisions regarding LLM migration. }
To summarize, in this paper, we make the following contributions:
\begin{itemize}
\item We propose \textit{ReCatcher}, the first systematic LLM regression testing framework for Python code generation across multiple code quality aspects.
\item We use our proposed framework to assess the impact of model updates across three key scenarios: fine-tuning, merging, and new model release.
\item We open-source \textit{ReCatcher}, providing a reusable and extensible regression testing framework to support informed LLM updating decisions~\cite{ReCatcherReplication}.
\rev{\item We provide a dataset of LLM-generated code snippets, covering nine model versions and two benchmarks available through Zenodo \cite{datasetZenodo}.}
\end{itemize}

The remainder of this paper is organized as follows: 
Section \ref{sec: backgound} provides the background.  
Section \ref{sec:inefficiency-tests} introduces the efficiency tests used for assessing LLM-generated Python code.  
Section \ref{sec: recatcher} details how \textit{ReCatcher} integrates these tests to systematically assess regressions, presenting its main concepts and workflow.  
Section \ref{sec: results} describes the experimental setup and analyzes the results, \rev{while Section \ref{sec:discussion} further discusses them and related implications.}
Section \ref{sec: related_work} discusses related work, followed by threats to validity in Section \ref{sec: threats_validity}.  
Finally, Section \ref{sec: conclusion} concludes the paper.  

\section{Background}
\label{sec: backgound}
In this section, we first introduce the selected LLMs for code generation. 
Then, we discuss the model adaptation techniques analyzed: fine-tuning and merging.

\subsection{LLMs for Code Generation}

\subsubsection{DeepSeek-Coder} is an open-source model family trained from scratch on 2 trillion tokens across 87 programming languages. 
Its training data is organized at the repository level to enhance code generation~\cite{guo2024deepseek}. 
The models utilize fill-in-the-middle and next-token prediction techniques and support 16K context window~\cite{guo2024deepseek}. 
Available sizes range from 1.3B to 33B parameters, with base and instruction-tuned versions~\cite{guo2024deepseek}. 

\subsubsection{CodeLlama} is an open-source model family based on Llama 2 from Meta~\cite{touvron2023llama} optimized for code completion and infilling.
The family includes three models: foundation, a Python-specialized, and an instruction-following, available in different sizes (7B, 13B, 34B, and 70B parameters). 
CodeLlama is primarily trained on a near-deduplicated dataset of publicly available code, with 8\% sourced from natural language data related to enhance comprehension~\cite{roziere2023code}. 

\subsubsection{GPT-Family} OpenAI models have demonstrated strong code generation capability~\cite{evalplusLeaderboard}. 
GPT-3.5, optimized for chat applications, includes the GPT-3.5-turbo model, which is fine-tuned for enhanced language comprehension and text generation capabilities. 
The GPT-4 family introduces GPT-4o, a multimodal model capable of processing and generating text, images, and audio. 
GPT-4o-mini is a smaller, more cost-efficient version of GPT-4o, offering similar capabilities at a reduced computational cost~\cite{achiam2023gpt, openai2024gpt4omini, openai2023gpt35}. 

\subsection{LLMs Model Adaptation Techniques} 
Model adaptation tailors LLMs for specific tasks or domains by leveraging their pre-existing capabilities, reducing the need for extensive training from scratch~\cite{ling2023domain}. 

\subsubsection{Merging} Model merging (or fusing) combines parameters from multiple LLMs to create a new model with integrated capabilities without requiring access to the original training data~\cite{dehghan2024mergerepair}. 
This technique preserves and enhances the strengths of each merged model. 
TIES (TrIm, Elect Sign, and Merge) is a notable merging method that: (i) resets minimally changed parameters during fine-tuning, (ii) resolves sign conflicts among parameters, and (iii) merges only aligned parameters to retain their strengths~\cite{yadav2024ties}. 
Merging has been applied to different SE tasks~\cite{dehghan2024mergerepair}.

\subsubsection{Fine-tuning} Fine-tuning adjusts LLM's parameters through additional training on a smaller, domain-specific dataset, enhancing its capability at addressing domain-specific tasks~\cite{lu2024fine}. 
Fine-tuning can be supervised and unsupervised, depending on data availability. 
Instruction fine-tuning further refines models by exposing them to structured prompts and expected outputs~\cite{parthasarathy2024ultimate}. 
In SE, LLMs have been fine-tuned to improve automated code review~\cite{yu2024fine} and code generation~\cite{tsai2024code}. 

\section{Automating LLMs Code Generation Capability Assessment}
\label{sec:inefficiency-tests}

To assess LLMs' code generation capability, we focus on the generated code, as a more capable LLM should generate more efficient code. 
This approach aligns with existing methods, such as CodeBLEU and pass@k~\cite{jiang2024survey}, which also evaluate LLMs based on generated code.
Directly assessing code efficiency is challenging due to its multi-faceted nature. 
To address this, we take an inverse approach, identifying inefficiencies as a proxy for efficiency, mirroring two sides of the same coin relationship between correctness and incorrectness~\cite{o2019incorrectness}. 
By systematically detecting inefficiencies, we provide a structured way to assess LLM-generated code beyond functional correctness, uncovering critical quality aspects that traditional evaluating approaches overlook. 

Prior studies have attempted to categorize LLM-generated code inefficiencies. 
For instance,~\cite{tambon2024bugs, dou2024s} focused on buggy code, grouping inefficiencies as underlying bug causes. 
However, inefficiencies are not limited to correctness errors. 
Moradi et al.~\cite{MORADIDAKHEL2023111734} highlighted that even syntactically and semantically correct code can still suffer from inefficiencies affecting readability, maintainability, and performance.
Building on this, our study takes a broader perspective on inefficiencies in LLM-generated code, considering issues regardless of correctness. 
To systematically assess these inefficiencies, we rely on a structured taxonomy~\cite{abbassi2025unveiling} serving as a foundation for developing dedicated efficiency tests, enabling a comprehensive evaluation of LLM-generated code quality.
The used taxonomy groups inefficiencies into five main categories \textit{General Logic}, \textit{Performance}, \textit{Maintainability}, \textit{Readability}, and \textit{Errors}, further divided into 19 sub-categories~\cite{abbassi2025unveiling}.
To ensure a structured assessment, we group these inefficiencies into three key aspects: logical correctness, static code issues, and performance.
Specifically, logical correctness is mapped to the \textit{General Logic} (\textit{Wrong Logic}, \textit{Partially Wrong Logic}, \textit{Wrong Method Input}, and \textit{Exception/Corner Case Handling}) category. Performance is mapped to the \textit{Performance} subcategories (\textit{Sub-optimal (Memory)} and \textit{Sub-optimal (Time)}). Static code issues include \textit{Errors} (Syntax Error, Missing Declaration/Import), \textit{Readability} (Confusing Variable Naming, Sub-readable Code), and \textit{Maintainability} (Code Duplication, Comment Duplication, Unnecessary Else, Unnecessary Conditional Block).

Inefficiencies are not unique to LLM-generated code, as human-written code is also susceptible to them~\cite{fry2012human, 10.1145/3196321.3196342}.
Consequently, significant efforts have been made to develop tests for inefficiency detection.
Below, we outline existing methods and discuss how we leveraged them to detect inefficiencies in LLM-generated code.

\subsection{Logical correctness}
According to the taxonomy of inefficiencies~\cite{abbassi2025unveiling}, \textit{General Logic} subcategories share a common characteristic—their impact on logical correctness.  
The automation of logical correctness assessment has been widely studied, with various approaches proposed in the literature. 
Formal verification-based methods are particularly notable, employing concepts such as state machines and symbolic execution~\cite{cotroneo2024automating}. 
However, such methods are inherently complex~\cite{faria2023case} and involve substantial computational overhead~\cite{tornblom2021scaling}.
With the emergence of LLMs, researchers have also leveraged LLMs themselves as automated judges to evaluate code correctness. 
However, an LLM’s ability to generate correct code does not necessarily mean it can accurately assess the correctness of other code~\cite{zhao2024codejudge}. 

Given these limitations, we opt for a more practical, stable and scalable approach: leveraging unit tests to automate logical correctness assessment. 
Unit tests execute the code with predefined inputs and compare the actual outputs to expected results. 
If a given code snippet passes all tests associated with its initial task, it is considered logically correct; otherwise, failures indicate inefficiencies in logic~\cite{10.5555/948785.948830}.
Despite potential challenges, such as incomplete coverage and weak test suites, unit tests remain an effective tool for validating code correctness, as they can be complemented by generated-tests using tools like EvoSuite~\cite{fraser2011evosuite} or by leveraging LLMs to enhance coverage~\cite{fraser2011evosuite, bhatia2024unit, ryan2024code}.
In our approach, we do not explicitly automate the detection of all inefficiency subcategories within \textit{General Logic}. 
Instead, we focus on detecting their consequences, namely through test failures.
This proxy-based approach offers a practical solution for assessing logical correctness at a category level in LLM-generated code. 

\subsection{Static Code Issues}
Static code issues are patterns or structures in code that negatively impact software quality across various aspects, including readability, maintainability, and performance~\cite{novak2010taxonomy, hassan2024evaluating}. 
Static analysis offers an effective approach for identifying and detecting such inefficiencies without requiring code execution~\cite{gomes2009overview}. 
This approach is fully automated, scalable, while helping developers catch issues in the development process and significantly reducing the time and effort needed for manual code review~\cite{singh2017evaluating}. 
Several static analysis tools are available, ranging from general-purpose tools that address common code issues to specialized tools focused on specific aspects of code, such as security vulnerability detection. 
These tools can be tailored to specific or support multiple programming languages.
For instance, SonarQube~\cite{sonarqube} provides seamless integration with CI/CD pipelines and supports a wide range of programming languages. 
Additionally, PMD, as a cross-programming language tool, includes Copy-Paste Detector (PMD-CPD) for identifying code duplication using configurable rule sets~\cite{pmd}.
However, such tools often introduce significant overhead compared to more specialized solutions, such as linters~\cite{habchi2018adopting}. 
Linters, typically designed for a single programming language, are easier to integrate into IDEs and allow for custom rule extensions~\cite{rafnsson2020fixing}.
To ensure scalability and effectiveness, we employ linters for detecting static code issues in LLM-generated code.  
Specifically, we use Pylint~\cite{pylint}, which is regarded as the most popular~\cite{gulabovska2019survey} and efficient Python linter~\cite{groce2021evaluating}. 
Pylint analyzes code by constructing an Abstract Syntax Tree (AST) and applying a set of predefined, extensible rules. Detected issues are reported and mapped to specific and unique predefined messages~\cite{pylint}.
To automate the detection of static inefficiencies from the taxonomy~\cite{abbassi2025unveiling}, we mapped each inefficiency subcategory to its corresponding Pylint messages by systematically running Pylint on a labeled inefficient code dataset~\cite{abbassi2025unveiling} and collecting its outputs.  
Most inefficiencies were mapped on a one-to-one or one-to-many basis with Pylint messages, where violations of these messages indicate inefficiencies.
However, certain inefficiencies required custom handling. 
For example, \textit{Missing Import} and \textit{Missing Variable Declaration} subcategories from the \textit{Error} category could not be distinguished as Pylint reports both under the same messages for undefined symbols. 
Consequently, we combined these two subcategories into a single test. 
Additionally, Pylint does not natively detect \textit{Unnecessary Conditional Block} inefficiency. 
To address this limitation, we extended Pylint by defining a custom rule that flags patterns matching \texttt{if condition: return True|False}.
Pylint also lacks native support for code duplication detection.  
Rather than extending Pylint for such a goal, we opted for an established external tool to ensure accuracy and reliability. 
PMD-CPD detects structural similarities using token-based approach allowing it to detect duplications even if they slightly differ in terms of variable names or formatting.
Given its effectiveness and reliability, we utilized PMD-CPD to automate the detection of code duplication and comment duplication, key subcategories in the taxonomy of inefficiencies~\cite{zakeri2023systematic, svajlenko2015evaluating}.

\subsection{Performance}
The \textit{Performance} category in the used taxonomy is organized into several subcategories at multiple levels~\cite{abbassi2025unveiling}. 
In our study, we focus on detecting inefficiencies at the top level, without further subdividing them into finer-grained subcategories. 
Several methods exist for automatically evaluating code performance. 
Static analysis, while applicable to performance assessment, occasionally yields incorrect results as performance inefficiencies are difficult to detect without execution, potentially leading to unnecessary code changes~\cite{cui2024empirical}.
Calotoiu et al.~\cite{calotoiu2013using} proposed a scalability bug detection approach, while various profiling techniques have emerged~\cite{bergel2012spy}. 
However, these methods are limited to assessing performance at the scalability level.
AST-based approaches, such as those used in LeetCode~\cite{nguyen2024development}, are common in educational settings, though they are less effective for real-world code assessment. 
Assessing the performance of LLM-generated code is challenging, as it requires determining whether a given solution is optimal in terms of execution time and memory usage. 
Since our primary goal is to compare the performance of generated code from two models and detect potential regressions, we adopt a direct performance efficiency comparison approach rather than evaluating a single solution in isolation.
Specifically, we execute the LLM-generated code from both models and profile their execution time and memory usage. We measure execution time by recording the start and end times.  
For memory usage, we leverage tarcemalloc from Python \cite{PythonTracemalloc}, allowing us to track the memory usage through execution.
By leveraging this direct comparison approach, we can effectively detect regressions, improvements, or stability in LLM-generated code snippets across execution time and memory usage. 

\section{ReCatcher: LLMs Regression Testing Framework for Code Generation}
\label{sec: recatcher}
In this section, we introduce ReCatcher, our LLMs regression testing framework for Python code generation.
We begin with an overview of the framework, and then we provide a detailed breakdown of its architecture.

\subsection{Overview} 
ReCatcher is an end-to-end systematic regression testing framework for Python code generation. 
ReCatcher takes as input two LLMs: a deployed, actively used model (LLM) and a potential replacement (LLM'), compares them, and generates a detailed regression report across various code aspects.
The framework evaluates the code generation based on a benchmark of programming tasks. 
ReCatcher follows a three-step process: 
\begin{itemize}
    \item Code Generation, where code snippets are generated for the code tasks reported in the benchmarks, using both LLMs with repeated generation (\textit{m} times) to mitigate randomness; 
    \item Test Execution, where efficiency tests are executed to assess logical correctness, static issues, and performance. To ensure statistical robustness, performance tests are repeated \textit{n} times; and 
    \item Analysis, where test outputs are used to generate a comprehensive comparative report comparing code generation capability between LLM and LLM', highlighting improvement and regression. 
\end{itemize} 

\subsection{Architecture} 
\label{sec:architecture}
\begin{figure*}[h!]
  \centering
  \includegraphics[width=0.9\linewidth]{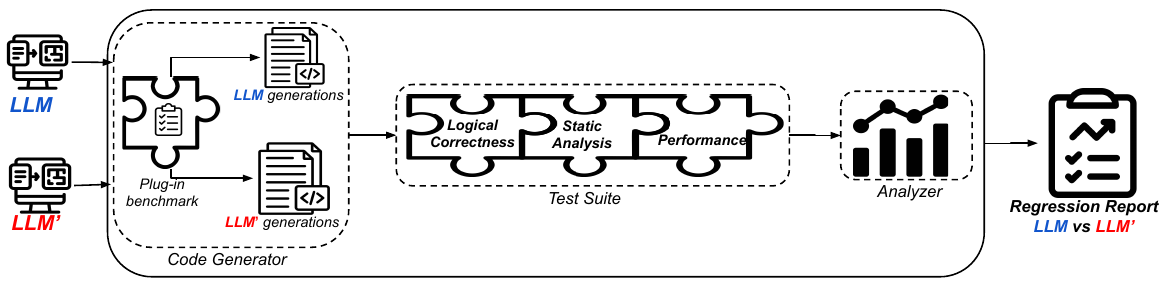}
  \caption{ReCatcher Architecture: LLM Regression Testing Framework for Code Generation}
\label{fig:recatcher:architecture+workflow}
\end{figure*}

Figure \ref{fig:recatcher:architecture+workflow} illustrates ReCatcher’s architecture, which follows a modular, plugin-based architecture providing flexibility and customization for testing LLMs. 
ReCatcher consists of three main components: Code Generator, Test Suite, and Analyzer. 
Below, we describe each component in detail. 
 
\subsubsection{Code Generator} This component is responsible for generating code using the two LLMs under test (LLM and LLM') based on a specified evaluation benchmark consisting of programming tasks. 
Since LLM-generated code can be non-deterministic, the generation process is repeated \texttt{m} times to mitigate randomness and ensure result consistency. 
The default repetition number is 10, however, users can adjust according to their available resources and evaluation needs. 

ReCatcher allows users to integrate different code generation benchmarks.
To ensure compatibility, the evaluation benchmark must meet two conditions: (i) be specifically designed for Python code generation tasks, as the framework currently focuses exclusively on Python; and (ii) include unit tests for each task, which are crucial for verifying logical correctness and assessing performance. 
To extend ReCatcher with a new benchmark, users must implement an interface that allows the framework to interact with the benchmark and extract unit tests. This integration ensures that ReCatcher can evaluate code generation against the newly added benchmark.
ReCatcher currently supports HumanEval+~\cite{liu2024your} and BigCodeBench~\cite{zhuo2024bigcodebench}, but additional benchmarks can be integrated. 
By supporting diverse evaluation scenarios, ReCatcher ensures meaningful, actionable results across various use cases, making it a flexible and extensible regression testing framework for LLM-generated code.
    
\subsubsection{Test Suite} This component is the core of ReCatcher, responsible for assessing LLMs' code generation capability using the tests defined in Section \ref{sec:inefficiency-tests}. 
The \textit{Test Suite} runs the tests on the code generated by \textit{Code Generator} across three code aspects: logical correctness, static code issues, and performance. The collected test outputs are then passed to the component, \textit{Analyzer}.

For logical correctness and static code issues, each test is conducted independently on every generated code snippet for each LLM, task, and repetition. 
Although unit tests, which assess logical correctness, can sometimes exhibit flaky behavior, we did not observe flakiness in our experiments. 
Therefore, each unit test is executed once per generated code snippet. 
As output, \textit{Test Suite} outputs a labeled dataset, where each snippet is annotated with detected inefficiencies. These results are aggregated to track inefficiency occurrences across repetitions and LLMs for further analysis by the \textit{Analyzer} component. 

For performance testing, only functionally correct code is considered. 
If a generated code is incorrect, its execution time or memory usage is irrelevant, as it does not fulfill the intended task and could introduce misleading conclusions; for example, an LLM generating a faster yet incorrect solution would not be a meaningful improvement.
To ensure fair and reliable comparisons, performance tests are conducted only when both LLMs generate correct code for a given benchmark task. 
To mitigate execution randomness, each test is repeated \texttt{n} times. 
The default test repetition is 5, however, users can adjust according to their available resources.
Performance evaluation supports two types of input data. If available, users can provide large inputs in JSONL format to test LLMs' performance under more demanding conditions. The \textit{Test Suite} executes the generated codes with large inputs, enabling performance assessment under more demanding conditions. 
If large inputs are not available, \textit{Test Suite} utilizes all inputs from existing unit tests to conduct performance testing.
In our study, large inputs were not available, so we conducted performance testing using inputs from unit tests provided with the code generation benchmark. 
\textit{Test Suite} generates a labeled dataset associating each task with correct outputs from both LLMs, with the distribution of execution time and memory usage per LLM. 
 
\textit{Test Suite} is highly customizable to accommodate different user needs: tests can be enabled or disabled based on the user's evaluation priorities or resource constraints. 
Additionally, \textit{Test Suite} is easily extensible, allowing users to integrate custom tests by providing a simple interface, ensuring smooth and consistent functionality within the ReCatcher framework.

\subsubsection{Analyzer} This component processes the detailed outputs from \textit{Test Suite} component and generates a comprehensive report comparing the code generation capability of the two models under evaluation (LLM and LLM'). 

For logical correctness and static code issues, the \textit{Analyzer} takes as input the number of inefficiencies detected per repetition for each model. 
The analysis begins by calculating the average number of detected inefficiencies per LLM, then quantifying the impact using the percentage:  
\begin{equation*}
    \scalebox{0.85}{$
    \text{Percentage Difference} = 
    \frac{
        \overline{\text{Inefficiencies}}_{\scriptstyle \text{LLM}} 
        - 
        \overline{\text{Inefficiencies}}_{\scriptstyle \text{LLM'}}
    }
    {\text{Total Tasks in Benchmark}} 
    \times 100
    $}
\end{equation*}
A positive value indicates that LLM' improves over LLM, whereas a negative value suggests that LLM' introduces regressions. 
The percentage differences are calculated per tested code efficiency, providing fine-grained insights into areas where improvements or regressions occur. 
For performance testing, the \textit{Analyzer} evaluates the execution time and memory usage distributions for each task to determine whether one LLM-generated code significantly outperforms the other. 
The task performance distribution is obtained from the \textit{Test Suite} component. 
Since we consider only functionally correct code snippets for a given task, the resulting samples are independent and non-paired. 
Under this condition, we apply the Mann-Whitney U test, a non-parametric statistical method designed for comparing two independent distributions. 
Unlike parametric tests, Mann-Whitney U does not assume normality, making it more appropriate for assessing differences in execution time and memory usage distributions between LLM-generated code. 
The Mann-Whitney U test evaluates the statistical significance of performance differences, determining whether an observed change indicates an improvement, regression, or no significant difference.
To ensure meaningful comparisons, we compute the ratio of tasks where LLM' exhibits improvements or regressions by normalizing against the total number of tasks in the benchmark. 
This normalization ensures that results remain interpretable across different benchmarks and task distributions.

ReCatcher’s final report includes two key metrics: (i) the inefficiency difference rate of inefficiency differences observed for logical correctness and static code analysis between LLM and LLM' per test; and (ii) the ratio of tasks exhibiting performance regressions or improvements based on execution time and memory usage. 
For users requiring more detailed and advanced analysis beyond aggregated metrics, the \textit{Test Suite} logs allow deeper investigation. 

\section{Experimental Setup and  Results}
\label{sec: results}
In this section, we present how we use ReCatcher (see Section \ref{sec: recatcher}) to assess regression in LLM-based code generation. 
We evaluate three update scenarios for models: 
\begin{itemize}
    \item \textbf{Scenario 1: Fine-tuning Impact} – Examining how fine-tuning affects code generation capabilities of LLMs.
    \item \textbf{Scenario 2: Merging Impact} – Examining how merging affects LLMs-based code generation. 
    \item \textbf{Scenario 3: Model Release Impact} – Evaluating regressions introduced by new versions of a model within a family of LLMs.
\end{itemize}

Initially, we detail the model and benchmark selection, then the experiment setup and potential regression in code generation capability. 

\subsection{Model Selection} 
In this work, we study regression across three update scenarios for models: fine-tuning, merging, and family model releases, for both open-source and proprietary models. 
For open-source models, we selected CodeLlama-7B and DeepSeek-Coder-6.7B as base models due to their widespread adoption and strong code generation capability on leaderboards among 7B-parameter models~\cite{evalplusLeaderboard}.
To effectively assess the influence of each scenario, we examined them separately. 
Specifically, to isolate the impact of fine-tuning and merging, we choose two variants per model (one fine-tuned and the other merged) from the Hugging Face Hub~\cite{HuggingFaceHub}. 
We prioritized variants that resulted from a single adaptation technique (either fine-tuning or merging), and we reviewed their details (e.g., fine-tuning datasets and merging methods) through model cards.
Selection was based on the highest number of downloads in Nov 2025 (last checked on Nov 27, 2024), with likes as a tiebreaker.
The fine-tuned variants were adapted using the Kotlin Exercises dataset~\cite{huggingface_jetbrains_kexercises}, while the merged variants included: (i) CodeLlama-7B merged with Llama2-7B, and (ii) DeepSeek-Coder-7B merged with DeepSeek-Coder Instruct and OpenCodeInterpreter-7B, both merged using the TIES 
technique~\cite{yadav2024ties}. 
For proprietary models, we selected GPT family models: GPT-3.5-turbo, GPT-4o, and GPT-4o-mini. 
When conducting this study, GPT-3.5 and GPT-4 represent the latest major releases, while GPT-4o-mini was included for its cost-efficiency and comparable performance to GPT-4o~\cite{evalplusLeaderboard}.

\subsection{Benchmark Selection}
We selected HumanEval+~\cite{liu2024your} and BigCodeBench~\cite{zhuo2024bigcodebench} as benchmarks to configure the Code Generator component in \textit{ReCatcher}. HumanEval+ is a widely used benchmark consisting of 164 manually crafted algorithmic tasks, each with unit tests for correctness assessment. 
While effective in evaluating code generation~\cite{jiang2024survey}, HumanEval+ is limited to algorithmic tasks and does not include real-world software development challenges. 
To address this limitation, we included BigCodeBench~\cite{zhuo2024bigcodebench}, which features 1,140 fine-grained tasks. The tasks require LLMs to invoke functions from 139 libraries across 7 different domains. 
Each task includes test cases, making it well-suited for effective evaluation of code generation. 
By combining HumanEval+ and BigCodeBench, we ensure a robust evaluation across both algorithmic and real-world programming tasks.

\subsection{Experimental Setup}
\label{sec:experimental_setup}
To systematically evaluate LLM capabilities through code inefficiencies, we designed a structured experimental setup covering code generation, test execution, and hardware configurations.

\subsubsection{Code-Generation Setup} When using the models to generate code for the benchmarks' tasks, we consider the following configurations:
\begin{itemize}
    \item \textbf{temperature = 0.1}, to provide more accurate and deterministic outputs with minimal randomness~\cite{lin2024llm};
    \item \textbf{max tokens = 2048}, aiming to accommodate sufficiently long code generations;
    \item \textbf{top-p = 0.95}, to balance diversity and coherence in generated outputs;
    \item \textbf{default prompt}, the method signature and task description from the benchmarks serve as input for the LLMs; 
    \item \textbf{model access}, GPT models were queried on January 18, 2025.
\end{itemize}

\subsubsection{ReCatcher Setup} Aiming to support the reproducibility of our results and accounting for variations in generation and execution, we applied multiple trials:
\begin{itemize}
    \item \textbf{Code Generation Repetitions (n=10)}, each prompt (task) is generated 10 times per model;
    \item \textbf{Performance Test Repetition (m=5)}, each generated code snippet is executed 5 times.
\end{itemize}

\subsection{Hardware Setup} To run our experiments, while aiming to reduce costs, we adopted different strategies, specifically for the tasks in BigCodeBench, as they require more powerful computational resources. Below, we detail the hardware setup in our study:
\begin{itemize}
    \item \textbf{Code Generation:} NVIDIA A100 (32GB);
    \item \textbf{Static Tests:} 12 vCPUs with 32GB RAM;
    \item \textbf{General Logic and Performance Tests:} NVIDIA V100 (32GB).
\end{itemize}

\subsection{Results}

Table \ref{tab: all_results_regression} summarizes the regression test results across the three update scenarios: fine-tuning, merging, and model release.
In the following, we analyze the results for each scenario, highlighting key trends, regressions, and improvements in code generation capability.   

\begin{table*}[]
    \centering
    \caption{Regression Testing Results Across Model Update Scenarios}
    \resizebox{\textwidth}{!}{ 
\begin{tabular}{ccc|cc|cc|cc|}
\cline{4-9}
                                                     &                                                       &                                                                              & \multicolumn{2}{c|}{CodeLlama}                                                                                                                       & \multicolumn{2}{c|}{DeepSeek-Coder}                                                                                                                  & \multicolumn{2}{c|}{GPT Family}                                                                                                              \\ \cline{4-9} 
                                                     &                                                       &                                                                              & \multicolumn{1}{c|}{\begin{tabular}[c]{@{}c@{}}Fintuned vs \\ Original\end{tabular}} & \begin{tabular}[c]{@{}c@{}}Merged vs \\ Original\end{tabular} & \multicolumn{1}{c|}{\begin{tabular}[c]{@{}c@{}}Fintuned vs \\ Original\end{tabular}} & \begin{tabular}[c]{@{}c@{}}Merged vs \\ Original\end{tabular} & \multicolumn{1}{c|}{\begin{tabular}[c]{@{}c@{}}GPT \\ 3.5 vs 4o\end{tabular}} & \begin{tabular}[c]{@{}c@{}}GPT \\ 4o vs 4o-mini\end{tabular} \\ \hline
\multicolumn{1}{|c|}{\multirow{15}{*}{HumanEval+}}    & \multicolumn{1}{c|}{General Logic}                    & Incorrect Code                                                               & \multicolumn{1}{c|}{1.95\%}                                                           & -18.72\%                                                       & \multicolumn{1}{c|}{14.94\%}                                                          & 25.98\%                                                        & \multicolumn{1}{c|}{4.15\%}                                                    & 3.48\%                                                        \\ \cline{2-9} 
\multicolumn{1}{|c|}{}                               & \multicolumn{1}{c|}{\multirow{2}{*}{Errros}}          & Syntax Error                                                                 & \multicolumn{1}{c|}{-12.93\%}                                                         & -9.02\%                                                        & \multicolumn{1}{c|}{-1.04\%}                                                          & 1.22\%                                                         & \multicolumn{1}{c|}{-0.12\%}                                                   & 0.55\%                                                        \\ \cline{3-9} 
\multicolumn{1}{|c|}{}                               & \multicolumn{1}{c|}{}                                 & \begin{tabular}[c]{@{}c@{}}Missing \\ Declaration/Import\end{tabular}        & \multicolumn{1}{c|}{-1.89\%}                                                          & 3.54\%                                                         & \multicolumn{1}{c|}{1.16\%}                                                           & 0.79\%                                                         & \multicolumn{1}{c|}{-10.73\%}                                                  & 10.55\%                                                       \\ \cline{2-9} 
\multicolumn{1}{|c|}{}                               & \multicolumn{1}{c|}{\multirow{4}{*}{Maintainability}} & Code Duplication                                                             & \multicolumn{1}{c|}{2.07\%}                                                           & -2.87\%                                                        & \multicolumn{1}{c|}{1.16\%}                                                           & 1.65\%                                                         & \multicolumn{1}{c|}{5.06\%}                                                    & -0.61\%                                                       \\ \cline{3-9} 
\multicolumn{1}{|c|}{}                               & \multicolumn{1}{c|}{}                                 & Comment Duplication                                                          & \multicolumn{1}{c|}{4.51\%}                                                           & -4.45\%                                                        & \multicolumn{1}{c|}{2.44\%}                                                           & 2.50\%                                                         & \multicolumn{1}{c|}{0.00\%}                                                    & 0.00\%                                                        \\ \cline{3-9} 
\multicolumn{1}{|c|}{}                               & \multicolumn{1}{c|}{}                                 & Unnecessary Else                                                             & \multicolumn{1}{c|}{16.22\%}                                                          & 7.80\%                                                         & \multicolumn{1}{c|}{7.87\%}                                                           & 7.44\%                                                         & \multicolumn{1}{c|}{4.39\%}                                                    & -0.24\%                                                       \\ \cline{3-9} 
\multicolumn{1}{|c|}{}                               & \multicolumn{1}{c|}{}                                 & \begin{tabular}[c]{@{}c@{}}Unnecessary \\ Conditional Block\end{tabular}      & \multicolumn{1}{c|}{0.98\%}                                                           & 1.89\%                                                         & \multicolumn{1}{c|}{0.67\%}                                                           & 1.10\%                                                         & \multicolumn{1}{c|}{0.85\%}                                                    & 0.18\%                                                        \\ \cline{2-9} 
\multicolumn{1}{|c|}{}                               & \multicolumn{1}{c|}{\multirow{2}{*}{Readability}}     & \begin{tabular}[c]{@{}c@{}}Confusing \\ Variable Naming\end{tabular}         & \multicolumn{1}{c|}{1.22\%}                                                           & -0.18\%                                                        & \multicolumn{1}{c|}{0.55\%}                                                           & -1.28\%                                                        & \multicolumn{1}{c|}{2.87\%}                                                    & 0.00\%                                                        \\ \cline{3-9} 
\multicolumn{1}{|c|}{}                               & \multicolumn{1}{c|}{}                                 & \begin{tabular}[c]{@{}c@{}}Sub-readable \\ Code\end{tabular}                 & \multicolumn{1}{c|}{1.59\%}                                                           & 1.22\%                                                         & \multicolumn{1}{c|}{2.93\%}                                                           & 2.38\%                                                         & \multicolumn{1}{c|}{0.67\%}                                                    & 0.00\%                                                        \\ \cline{2-9} 
\multicolumn{1}{|c|}{}                               & \multicolumn{1}{c|}{\multirow{6}{*}{Performance}}      & \begin{tabular}[c]{@{}c@{}}Memory Usage:\\ Improvement\end{tabular} & \multicolumn{1}{c|}{0.61\%}                                                           & 0.00\%                                                         & \multicolumn{1}{c|}{0.61\%}                                                           & 0.00\%                                                         & \multicolumn{1}{c|}{0.61\%}                                                    & 0.00\%                                                        \\ \cline{3-9} 
\multicolumn{1}{|c|}{}                               & \multicolumn{1}{c|}{}                                 & \begin{tabular}[c]{@{}c@{}}Memory Usage:\\ Regression\end{tabular}           & \multicolumn{1}{c|}{0.00\%}                                                           & 0.00\%                                                         & \multicolumn{1}{c|}{1.83\%}                                                           & 1.22\%                                                         & \multicolumn{1}{c|}{0.00\%}                                                    & 0.61\%                                                        \\ \cline{3-9} 
\multicolumn{1}{|c|}{}                               & \multicolumn{1}{c|}{}                                            & \begin{tabular}[c]{@{}c@{}}Execution Time\\ Improvement\end{tabular}         & \multicolumn{1}{c|}{26.22\%}                                                          & 10.98\%                                                        & \multicolumn{1}{c|}{46.34\%}                                                          & 48.17\%                                                        & \multicolumn{1}{c|}{65.24\%}                                                   & 0.61\%                                                        \\ \cline{3-9} 
\multicolumn{1}{|c|}{}                               & \multicolumn{1}{c|}{}                                 & \begin{tabular}[c]{@{}c@{}}Execution Time\\ Regression\end{tabular}          & \multicolumn{1}{c|}{0.61\%}                                                           & 0.00\%                                                         & \multicolumn{1}{c|}{0.61\%}                                                           & 1.22\%                                                         & \multicolumn{1}{c|}{0.00\%}                                                    & 80.49\%                                                       \\ \hline
\multicolumn{1}{|c|}{\multirow{15}{*}{BigCodeBench}} & \multicolumn{1}{c|}{General Logic}                    & Incorrect Code                                                               & \multicolumn{1}{c|}{3.41\%}                                                           & 6.71\%                                                         & \multicolumn{1}{c|}{9.32\%}                                                           & 1.95\%                                                         & \multicolumn{1}{c|}{-12.19\%}                                                  & 11.39\%                                                       \\ \cline{2-9} 
\multicolumn{1}{|c|}{}                               & \multicolumn{1}{c|}{\multirow{2}{*}{Errros}}          & Syntax Error                                                                 & \multicolumn{1}{c|}{-0.69\%}                                                          & -30.77\%                                                       & \multicolumn{1}{c|}{-3.44\%}                                                          & 0.26\%                                                         & \multicolumn{1}{c|}{-2.62\%}                                                   & -0.25\%                                                       \\ \cline{3-9} 
\multicolumn{1}{|c|}{}                               & \multicolumn{1}{c|}{}                                 & \begin{tabular}[c]{@{}c@{}}Missing \\ Declaration/Import\end{tabular}        & \multicolumn{1}{c|}{3.38\%}                                                           & 2.55\%                                                         & \multicolumn{1}{c|}{1.65\%}                                                           & 1.54\%                                                         & \multicolumn{1}{c|}{-51.64\%}                                                  & 28.31\%                                                       \\ \cline{2-9} 
\multicolumn{1}{|c|}{}                               & \multicolumn{1}{c|}{\multirow{4}{*}{Maintainability}} & Code Duplication                                                             & \multicolumn{1}{c|}{3.47\%}                                                           & -20.36\%                                                       & \multicolumn{1}{c|}{1.49\%}                                                           & 0.72\%                                                         & \multicolumn{1}{c|}{1.13\%}                                                    & -1.18\%                                                       \\ \cline{3-9} 
\multicolumn{1}{|c|}{}                               & \multicolumn{1}{c|}{}                                 & Comment Duplication                                                          & \multicolumn{1}{c|}{10.16\%}                                                          & 1.39\%                                                         & \multicolumn{1}{c|}{0.39\%}                                                           & 0.40\%                                                         & \multicolumn{1}{c|}{0.03\%}                                                    & 0.00\%                                                        \\ \cline{3-9} 
\multicolumn{1}{|c|}{}                               & \multicolumn{1}{c|}{}                                 & Unnecessary Else                                                             & \multicolumn{1}{c|}{1.90\%}                                                           & -1.90\%                                                        & \multicolumn{1}{c|}{0.10\%}                                                           & 0.08\%                                                         & \multicolumn{1}{c|}{1.07\%}                                                    & 0.78\%                                                        \\ \cline{3-9} 
\multicolumn{1}{|c|}{}                               & \multicolumn{1}{c|}{}                                 & \begin{tabular}[c]{@{}c@{}}Unnecessary \\ Conditional Block\end{tabular}     & \multicolumn{1}{c|}{-0.69\%}                                                          & -0.21\%                                                        & \multicolumn{1}{c|}{-0.59\%}                                                          & 0.01\%                                                         & \multicolumn{1}{c|}{0.15\%}                                                    & -0.01\%                                                       \\ \cline{2-9} 
\multicolumn{1}{|c|}{}                               & \multicolumn{1}{c|}{\multirow{2}{*}{Readability}}     & \begin{tabular}[c]{@{}c@{}}Confusing \\ Variable Naming\end{tabular}         & \multicolumn{1}{c|}{-1.24\%}                                                          & 3.60\%                                                         & \multicolumn{1}{c|}{0.40\%}                                                           & 0.31\%                                                         & \multicolumn{1}{c|}{0.88\%}                                                    & -0.23\%                                                       \\ \cline{3-9} 
\multicolumn{1}{|c|}{}                               & \multicolumn{1}{c|}{}                                 & \begin{tabular}[c]{@{}c@{}}Sub-readable \\ Code\end{tabular}                 & \multicolumn{1}{c|}{-0.07\%}                                                          & -0.14\%                                                        & \multicolumn{1}{c|}{0.44\%}                                                           & 0.30\%                                                         & \multicolumn{1}{c|}{0.57\%}                                                    & -0.02\%                                                       \\ \cline{2-9} 
\multicolumn{1}{|c|}{}                               & \multicolumn{1}{c|}{\multirow{6}{*}{Performance}}       & \begin{tabular}[c]{@{}c@{}}Memory Usage:\\ Improvement\end{tabular}          & \multicolumn{1}{c|}{0.35\%}                                                           & 2.28\%                                                         & \multicolumn{1}{c|}{1.23\%}                                                           & 2.72\%                                                         & \multicolumn{1}{c|}{0.96\%}                                                    & 0.61\%                                                        \\ \cline{3-9} 
\multicolumn{1}{|c|}{}                               & \multicolumn{1}{c|}{}                                 & \begin{tabular}[c]{@{}c@{}}Memory Usage:\\ Regression\end{tabular}           & \multicolumn{1}{c|}{0.61\%}                                                           & 1.49\%                                                         & \multicolumn{1}{c|}{0.61\%}                                                           & 9.04\%                                                         & \multicolumn{1}{c|}{0.88\%}                                                    & 1.32\%                                                        \\ \cline{3-9} 
\multicolumn{1}{|c|}{}                               & \multicolumn{1}{c|}{}                                       & \begin{tabular}[c]{@{}c@{}}Execution Time\\ Improvement\end{tabular}         & \multicolumn{1}{c|}{9.74\%}                                                           & 26.84\%                                                        & \multicolumn{1}{c|}{32.54\%}                                                          & 34.82\%                                                        & \multicolumn{1}{c|}{10.44\%}                                                   & 21.14\%                                                       \\ \cline{3-9} 
\multicolumn{1}{|c|}{}                               & \multicolumn{1}{c|}{}                                 & \begin{tabular}[c]{@{}c@{}}Execution Time\\ Regression\end{tabular}          & \multicolumn{1}{c|}{11.93\%}                                                          & 1.32\%                                                         & \multicolumn{1}{c|}{8.16\%}                                                           & 5.53\%                                                         & \multicolumn{1}{c|}{20.00\%}                                                   & 15.35\%                                                       \\ \hline
\end{tabular}
}
\label{tab: all_results_regression}
\end{table*}

\subsubsection{Scenario 1: Fine-tuning Impact}
For evaluating the impact of fine-tuning, we used (i) a fine-tuned variant of CodeLlama and (ii) a fine-tuned variant of DeepSeek-Coder, both fine-tuned on the same dataset, Kotlin Exercises~\cite{huggingface_jetbrains_kexercises}.
The most significant regression between the original and fine-tuned variant is observed in \textit{Syntax Error}. 
For example, the fine-tuned variant of CodeLlama, when tested on HumanEval+, showed 12.93\% regression in \textit{Syntax Error}, indicating a decline in capability to generate syntactically correct Python code. 
This regression, although varying in intensity, is present across the two used benchmarks and also on the fine-tuned variant of DeepSeek-Coder. 
A possible explanation is the nature of the fine-tuning dataset, which consists of Kotlin code, whereas the evaluation is performed on Python code.  
Given the significant syntactic differences between Kotlin and Python, fine-tuning may have disrupted the model's learned syntax, leading to more frequent syntax errors.
However, for \textit{Missing Declaration/Import}, although related to programming language, we did not notice regression (only fine-tuned CodeLlama on HumanEval+, which showed only 1.89\% regression); instead, we noticed a minor improvement in handling missing imports with a maximum of 3.38\%. 
These results suggest that while \textit{Syntax Error} were heavily impacted by language differences, \textit{Missing Declaration/Import} remained relatively stable, likely because such errors are linked to semantic understanding rather than strict syntax rules.

Despite the regression in \textit{Syntax Error}, we observed an overall improvement in \textit{General Logic}.
The improvement regarding \textit{General Logic} may be greater than the reported values since some logically correct implementations could not be evaluated due to syntax-related execution failures. 
Regarding \textit{Maintainability}, fine-tuned variants showed improvement, particularly in \textit{Comment Duplication}.
The CodeLlama variant improved by 10.16\% in \textit{Comment Duplication} on BigCodeBench, while DeepSeek-Coder showed an improvement, though to a lesser extent. 
One possible reason for this improvement is that DeepSeek-Coder was trained from scratch on a large-scale code dataset, making it better at handling redundancy and duplication.
Additionally, we observed an improvement in \textit{Unnecessary Else} statements, by 16.22\% and 7.87\%, for the fine-tuned variant of CodeLlama and DeepSeek-Coder, respectively. 
Such an improvement suggests that fine-tuning, even on a different programming language, may help LLMs generate more concise and structured conditional statements, improving maintainability.

The impact of fine-tuning on \textit{Readability} was limited, with percentage difference ranging between -1.24\% and 2.93\% across models.
In terms of memory usage, we observed no significant difference across models. 
However, for execution time, fine-tuning had mixed effects. 
The DeepSeek-Coder fine-tuned model on BigCodeBench showed a 46.34\% improvement in execution time, whereas the fine-tuned CodeLlama model exhibited a slight 2\% net regression (+9.74\% and -11.93\%, for improvement and regression, respectively).

Our results suggest that the impact of fine-tuning remains consistent when the same dataset is used for fine-tuning. 
However, when fine-tuning is performed on coding tasks from different programming languages, it can enhance logical reasoning but may also introduce syntax errors.

\subsubsection{Scenario 2: Merging Impact}
For evaluating the merging adaptation, we considered two merged variants: (i) CodeLlama merged with Llama2 and (ii) DeepSeek-Coder merged with OpenCodeInterpreter~\cite{zheng2024opencodeinterpreter}. 

The CodeLlama merged variant exhibited a regression of 30.77\% in \textit{Syntax Error} on BigCodeBench and 18.72\% on HumanEval+. 
This regression may be due to the merging process incorporating a general-purpose model (Llama2) not optimized for code generation, leading to syntax inconsistencies. 
The regression on \textit{Syntax Error} may explain the regression of 18.72\% in logical correctness on HumanEval+. 
However \textit{General Logic} improved by 6.71\% on BigCodeBench.
The actual improvement might be higher but we cannot precisely quantify it, because of syntax errors.
The merging impact difference between benchmarks may indicate that \textit{General Logic} is more affected in algorithmic-style tasks (as seen in HumanEval+) rather than in real-world programming tasks spanning diverse domains such as machine learning and data analytics (as covered by BigCodeBench). 
Regarding \textit{Missing Declaration/Import}, we observe an improvement of 3.54\% when tested on HumanEval+, suggesting that merging impact on \textit{Syntax Error} is more pronounced than on \textit{Missing Declaration/Import}. 

For \textit{Maintainability}, we observed a significant regression of 20.36\% in \textit{Code Duplication} on BigCodeBench. This regression suggests that merging with general-purpose LLM, which lacks code-specific optimization, may lead to inconsistent coding patterns, increasing redundancy.
Regarding \textit{Readability}, the impact remained minimal with differences rates ranging from 3.60\% to -0.18\%, highlighting that merging influences the syntax structure more than the overall readability. 
For \textit{Performance}, memory usage remained unchanged, while execution time improved by 26.84\% on BigCodeBench and 10.98\% on HumanEval+. 

Regarding the DeepSeek-Coder merged model, we observed overall improvements across all code aspects except for a minor minor exception regression regression of 1.28\% in \textit{Confusing Variable Naming} on HumanEval+. 
The improvements were overall more pronounced on HumanEval+, while BigCodeBench shows overall marginal improvements, possibly because the merged model (OpenCodeInterpreter) was trained on LeetCode-style programmatic tasks~\cite{zheng2024opencodeinterpreter}, which resembles HumanEval+.  
\textit{General Logic} showed improvement of 25.98\% on HumanEval+ and only 2\% on BigCodeBench.
\textit{Performance} showed a notable improvement of 48.18\% on HumanEval+ and 34.82\% on BigCodeBench regarding execution time. 
This improvement may suggest that merging with a model trained on LeetCode-style tasks emphasizes performance optimization, as these tasks prioritize algorithmic efficiency, minimal computation, and optimized execution paths. 
However, memory usage showed a net regression of approximately 6\% on BigCodeBench and only ~1\% HumanEval+. 
For \textit{Unnecessary Else}, we observed an improvement of 7.87\% when tested on HumanEval+, while results remain stable on BigCodeBench.
For the other code aspects, \textit{Readability}, \textit{Maintainability} and \textit{Errors}, we have seen minimal variations with differences in rates lower than 2\%. 

These findings emphasize that model selection plays a crucial role in merging, as it can either mitigate or amplify regressions in code generation capabilities.

\subsubsection{Scenario 3: Model Release Impact}
To assess the impact of model release, we studied two release scenarios: (i) migration from GPT-3.5-turbo to GPT-4o and (ii) migration from GPT-4o to GPT-4o-mini.
 
\subsubsection*{Experiment 1: Migration from GPT-3.5-turbo to GPT-4o} GPT-4o exhibited a significant regression in \textit{Missing Declaration/Import} of 10.73\% on HumanEval+ and 51.64\% on BigCodeBench. 
The regression is more pronounced in BigCodeBench, which can be attributed to its greater diversity across tasks beyond algorithmic problem-solving, requiring a wider range of libraries. 
Prior studies have also highlighted GPT models’ challenges in handling external libraries and issues with deprecated or incorrectly referenced methods, further reinforcing this observation~\cite{wang2024and}. 
We also observed a small regression in \textit{Syntax Errors} of  2.62\% on BigCodeBench. 
Both \textit{Missing Declaration/Import} and \textit{Syntax Error} directly impact the assessment of \textit{General Logic} by preventing the execution of test cases, as these errors cause the program to fail during runtime.
The combined effect of these regressions in \textit{Missing Declaration/Import} and \textit{Syntax Error} contributed to the 12.19\% regression in \textit{General Logic} on BigCodeBench.
Interestingly, although HumanEval+ exhibited a 10.73\% regression in \textit{Missing Declaration/Import}, it still showed a 4.15\% improvement in \textit{General Logic}. 
Regarding \textit{Performance}, we observed a notable improvement in execution time efficiency by 65.24\% on HumanEval+ and a net regression of approximately 10\% on BigCodeBench. 
Finally, GPT-4o demonstrated a slight improvement in \textit{Maintainability}, of 5.06\% in \textit{Code Duplication}, suggesting a minor structural improvement in generated code despite other regressions.

Although GPT-4o exhibited regressions across multiple aspects, it is still widely considered more capable than GPT-3.5-turbo. 
These findings highlight the importance of a comprehensive regression testing framework that evaluates models beyond predefined correctness benchmarks, ensuring a well-rounded assessment of code quality and performance.

\subsubsection*{Experiment 2: Migration from GPT-4o to GPT-4o-mini} 
GPT-4o-mini demonstrated minimal variations when compared with GPT-4o. 
\textit{Readability}, \textit{Maintainability}, and \textit{Syntax Error} remained largely unchanged, indicating that the smaller model preserved many of GPT-4o’s code generation capabilities.
Despite being a smaller and more cost-effective model, GPT-4o-mini outperforms GPT-4o, showing an 11.39\% improvement in \textit{General Logic} on BigCodeBench. 
Additionally, its handling of \textit{Missing Declaration/Import} improved by up to 28.31\% on BigCodeBench. 
However, the most significant regression was observed in execution time. 
GPT-4o-mini exhibited an 80.49\% regression on HumanEval+, while showing a slight improvement of 5\% on BigCodeBench. 
This regression indicates that the execution time of code snippets generated by GPT-4o-mini increased significantly for algorithmic tasks that require optimization, such as those in HumanEval+. 


\subsection{Comparison with Baseline Solutions}

\begin{table}[]
\caption{Regression Testing Results using LLM as a Judge and Accuracy Analysis}
\label{tab:regression_baseline}
\begin{tabular}{c|r|rr|}
\cline{2-4}
\multicolumn{1}{l|}{\multirow{2}{*}{}}                                                  & \multicolumn{1}{c|}{\multirow{2}{*}{Logical Correctness}} & \multicolumn{2}{c|}{Performance}                                                                                                                                    \\ \cline{3-4} 
\multicolumn{1}{l|}{}                                                                   & \multicolumn{1}{c|}{}                                     & \multicolumn{1}{c|}{\begin{tabular}[c]{@{}c@{}}Time \\ Execution\end{tabular}} & \multicolumn{1}{c|}{\begin{tabular}[c]{@{}c@{}}Memory \\ Consumption\end{tabular}} \\ \hline
\multicolumn{1}{|c|}{\begin{tabular}[c]{@{}c@{}}Regression/\\ Improvement\end{tabular}} & -3.01                                                     & \multicolumn{1}{r|}{7.80}                                                      & -1.45                                                                              \\ \hline
\multicolumn{1}{|c|}{Accuracy}                                                          & 0.80                                                      & \multicolumn{1}{r|}{0.58}                                                      & 0.66                                                                               \\ \hline
\multicolumn{1}{|c|}{Precision}                                                         & 0.88                                                      & \multicolumn{1}{r|}{0.18}                                                      & 0.19                                                                               \\ \hline
\multicolumn{1}{|c|}{Recall}                                                            & 0.40                                                      & \multicolumn{1}{r|}{0.13}                                                       & 0.13                                                                               \\ \hline
\multicolumn{1}{|c|}{F1-score}                                                          & 0.55                                                      & \multicolumn{1}{r|}{0.15}                                                      & 0.15                                                                                \\ \hline
\end{tabular}
\end{table}


\rev{To evaluate our framework in comparison with existing solutions, this section presents a focused analysis involving a subset of previously studied models and versions. Rather than repeating the full evaluation, we selected a representative configuration to enable a controlled and meaningful comparison with baseline methods. Many model configurations differ substantially in the quality aspects being measured, which makes them less suitable for fine-grained comparison. By selecting configurations with only minor quality differences, we intentionally stress-test both our framework and the baselines in their ability to detect subtle issues—precisely where evaluation methods tend to struggle.
This targeted subset of models allows us to assess how effectively baseline approaches capture nuanced quality variations that our framework is specifically designed to identify.}

\rev{
We apply the following criteria to select these evaluation models. First, based on the results in Table~\ref{tab: all_results_regression}, we identify pairs of models with minimal variation in logical correctness—specifically, a 1.95\% difference observed between (i) CodeLlama and its finetuned version on HumanEval+, and (ii) DeepSeek-Coder and its merged version on BigCodeBench.
Such marginal differences offer challenging test cases: if a baseline method can consistently detect and reflect these fine-grained distinctions, it demonstrates high sensitivity; if not, it reveals limitations in relying on LLMs for evaluation.
Following the same reasoning, we also consider performance-related results (Table~\ref{tab: all_results_regression}), which further motivate our selection of CodeLlama and its finetuned version on HumanEval+, as it reports smaller differences when comparing with DeepSeek-Coder and its merged version.
This setup enables a focused stress-test of baseline methods, emphasizing their ability to capture nuanced differences rather than trivially large gaps.}


\rev{Regarding the selection of baselines, we define them based on the different dimensions under evaluation. For \textit{Logical Correctness} and \textit{Performance}, we leverage the capabilities of large language models (LLMs) as automated judges—an approach motivated by recent studies employing LLMs for evaluation tasks 
\cite{zheng2023judging, gu2024survey}.
Such an approach enables scalable and reproducible evaluation beyond traditional syntactic or manual analysis. 
While the use of LLMs as judges is an emerging trend, our application to both dimensions under evaluation within our framework represents a novel integration, contributing to more nuanced and automated assessments. 
Moreover, we employ LLMs not just for correctness judgments, but also to reason about performance aspects. 
This multi-faceted use of LLMs as evaluators offers a more comprehensive and nuanced assessment of code quality than prior work focused on correctness alone.
For \textit{Static Code Issues}, we employ well-established software engineering metrics, providing a reliable and interpretable foundation for analysis. 
While these metrics themselves are standard, we use them to examine whether traditional static analysis can detect the same inefficiencies surfaced by ReCatcher. 
Table~\ref{tab:regression_baseline} summarizes the results discussed herein.}

\textbf{\textit{Logical Correctness}}. \rev{To evaluate such an aspect of the generated code snippets, we explored LLMs as automated judges, as previously stated.
Specifically, we used \texttt{GPT-4o-mini}, prompting it with a code snippet and its corresponding test cases, and asking whether the implementation was correct.
To promote transparency and reproducibility, we also requested an explanation for each judgment. 
Since we generated code snippets for both the original and fine-tuned versions of the model, repeating the process ten times (see Section~\ref{sec:experimental_setup}), the experiment resulted in 3,280 code snippets. 
In total, we obtained judgments for 3,226 snippets, with the remaining 54 cases discarded due to unhandled or unexpected outputs from the LLM.
Since some of the repeated generations produced identical or highly similar code snippets, we opted to run the inference only once per prompt for this evaluation. 
This decision was supported by the fact that variability across different runs was already captured in the broader set of ten repetitions performed during the initial code generation phase.  
To assess regression between the two model versions, we compared the total number of snippets classified as \textit{not correct}.
Additionally, since we had access to the test execution results for each code snippet (see Section~\ref{sec:experimental_setup}), we used these outcomes as ground truth, enabling us to measure the accuracy of using LLMs to evaluate the correctness of the generated code.
The scripts used for this evaluation, including the prompts submitted to the LLMs and the resulting outputs, are available in our online appendix~\cite{ReCatcherReplication}.}

\rev{Comparing the two model versions, we observed a regression of 3.01\% in logical correctness from the original to the finetuned model.
This contrasts with the improvement of 1.95\% reported by ReCatcher (see third row, Table~\ref{tab: all_results_regression}).
Checking the accuracy of using LLMs as a judge, we found that LLM-based judgments achieved an overall accuracy of 0.80 and a precision of 0.88.
However, there are some wrong judgments.
For example, consider the code snippet generated for the \texttt{HumanEval/7} task by CodeLlama (original version) presented in Figure \ref{code:llama-original}.
Overall, the code is responsible for checking whether the strings of a list contain a given substring, returning only the ones that apply. 
While all test cases report successful status, the LLM considered that such an implementation is wrong, emphasizing one specific assertion from the set of tests: \textit{an assertion that expects `grunt' and `prune' to be returned when filtering with the substring `run'. Neither `grunt' nor `prune' contains the substring `run', which makes this assertion incorrect.}
We can observe that the LLM's judgment is wrong, as both mentioned words, `grunt' and `prune', contain the given substring `run'.
}

\begin{figure}
    \centering
    \begin{lstlisting}[language=python, numbers=left]
from typing import List


def filter_by_substring(strings: List[str], substring: str) -> List[str]:
    """ Filter an input list of strings only for ones that contain given substring """
    return [s for s in strings if substring in s]

def check(candidate):
    assert candidate(['grunt', 'trumpet', 'prune', 'gruesome'], 'run') == ['grunt', 'prune']

check(filter_by_substring)
    \end{lstlisting}
    \caption{Code snippet generated by CodeLlama (Original version).}
    \label{code:llama-original}
\end{figure}


\rev{Moving forward, we observed that the recall was relatively low at 0.40, indicating that the LLM often failed to identify all incorrect implementations. 
This suggests a tendency toward false negatives, where incorrect code is mistakenly judged as correct.
For example, consider the code snippet generated for the \texttt{HumanEval/14} task by the CodeLlama, finetuned version, presented in Figure \ref{code:llama-finetuned}.
Such an implementation is expected to return a list of all non-empty prefixes of a given string, ordered from shortest to longest.
However, upon inspecting the implementation, we observe an error that causes the output to include an empty string at the beginning.
This occurs because in line 7, the loop starts with \texttt{i = 0}, and \texttt{string[:0]} evaluates to an empty string (\texttt{`'}), which gets appended as the first item in the prefix list.
The associated test cases are accurate to catch this error in the implementation, while the LLM erroneously reported that the implementation was correct, emphasizing that: \textit{The implementation iterates through the range of the string's length plus one, appending each prefix (from the start of the string to the current index) to the 'prefixes' list. The provided test assertions confirm that the function behaves as expected for various input cases [...]. Therefore, the implementation is correct.}
We also observed cases where the code snippets contained syntax errors that were completely overlooked by the LLM, as it focused solely on the implementation logic without properly detecting these issues during evaluation.
}


\begin{figure}
    \centering
    \begin{lstlisting}[language=python, numbers=left]
from typing import List


def all_prefixes(string: str) -> List[str]:
    """ Return list of all prefixes from shortest to longest of the input string """
    prefixes = []
    for i in range(len(string) + 1):
        prefixes.append(string[:i])
    return prefixes

def check(candidate):
    assert candidate('') == []

check(all_prefixes)
    \end{lstlisting}
    \caption{Code snippet generated by CodeLlama (Finetuned version).}
    \label{code:llama-finetuned}
\end{figure}

\textbf{\textit{Performance.}}
\rev{To evaluate this aspect, we also explored the use of LLMs as judges, using \texttt{GPT-4o-mini}. 
However, unlike the previous evaluation—where all code snippets were assessed—here we focus exclusively on code snippets deemed correct (see Section~\ref{sec:experimental_setup}), since our goal is to evaluate the LLM's ability to identify the better-performing implementation. 
This way, considering incorrect answers could introduce bias in our analysis.
We filtered and grouped the correct code snippets from both the original and finetuned models, resulting in 318 pairs of comparable solutions. 
Each pair was submitted to \texttt{GPT-4o-mini}, which was asked to determine which implementation performed better in terms of \textit{execution time} and \textit{memory consumption}.
We also requested that the model provide a rationale for each decision.
To assess regression between the two model versions, we counted how often each version was selected as the better option across the evaluated aspects. 
We repeated this process five times to account for the inherent randomness in LLM outputs, ensuring that the results were not biased by a single inference. 
For each repetition, we re-prompted the LLM with the same input pairs and collected its judgments and rationales independently. 
This allowed us to assess the stability of the LLM’s decisions across multiple runs.
Finally, to measure the accuracy of the LLM’s decisions, we executed each code snippet ten times, just like our original study, recording both execution time and memory usage.
To establish the ground truth for each pair, we applied the \textit{Mann-Whitney U} test to compare the distributions of performance metrics (e.g., execution time and memory consumption). Based on the statistical outcomes, we determined whether a significant difference existed. These results served as the reference to evaluate the accuracy of the LLM’s judgments. 
Again, the scripts used for this evaluation, including the prompts submitted to the LLMs and the resulting outputs, are available in our online appendix~\cite{ReCatcherReplication}.}

\rev{For execution time, we observed an average improvement of 7.8\% when comparing the original to the fine-tuned versions. 
This result aligns with ReCatcher’s findings (see Table \ref{tab: all_results_regression}), although the magnitude of improvement is significantly lower. 
This discrepancy may be attributed to the low F1-score of 0.15, which indicates poor overall model performance across all classes. Specifically, despite a moderate accuracy of 58\%, the macro-averaged precision (0.18) and recall (0.13) reveal that the model struggles to correctly identify and classify instances of less frequent classes. 
Such limitations may reduce the effectiveness of the fine-tuned models in more diverse or imbalanced settings, thereby explaining the more modest gains in execution time.
Turning to memory consumption, the LLM-based judgment indicated a regression of 1.45\%, whereas ReCatcher showed an improvement of 0.61\%, reflecting an opposite trend (see Table \ref{tab: all_results_regression}). 
Although the results differ, the interval among them is the lowest we observed in our analysis.
Moving forward with our analysis, 
this time, we observe a better accuracy for the LLM’s judgment (0.66) when compared to the time execution judgment.
However, the macro-averaged precision (0.19) and recall (0.13) remain low, leading to a modest F1-score of 0.15. 
These results reinforce the limitations of LLMs in consistently judging performance aspects across all classes, particularly when dealing with imbalanced or nuanced scenarios.
}

\rev{
During our analysis, we noticed that in some cases, the models generated identical or nearly identical code snippets, differing only in minor aspects such as extra blank spaces or variable names.
While LLMs correctly identified these cases as functionally equivalent and reported no difference in execution time or memory usage, ReCatcher proceeded to execute both snippets and collect runtime results.
However, we also found that even identical code snippets sometimes led to different performance metrics, introducing false positives in our results.
To mitigate this issue, we introduced a verification step before execution: we convert both code snippets into their respective Abstract Syntax Trees (ASTs) and compare them structurally.
If the ASTs are equivalent, we assume the code snippets have the same execution time and memory consumption, skipping redundant evaluation. Otherwise, we proceed with execution and comparison.
However, the results remained essentially unchanged, indicating that this potential threat does not significantly impact our findings.
}

\rev{These findings highlight that while LLMs performed reasonably well in assessing the \textit{logical correctness} of code snippets, they were notably less reliable when evaluating \textit{performance-related} aspects, such as execution time and memory usage, as evaluated here.
}

\textbf{\textit{Static Code Issues.}} 
\rev{To evaluate this aspect, we rely on standard software engineering metrics.
Specifically, we use Cyclomatic Complexity (CC) and the Maintainability Index (MI) \cite{coleman1994using}, as these metrics help practitioners estimate the structural complexity of code and the effort required to maintain it.
Following the same evaluation protocol described in Section~\ref{sec:architecture}, we analyze all generated code snippets, totaling 3,280 samples (1,640 from each model version).
Each snippet is temporarily saved to a local file and analyzed using the \texttt{radon} tool to extract the corresponding CC and MI values.\footnote{https://pypi.org/project/radon/}
To assess regression between the two model versions, we apply the Wilcoxon signed-rank test.
For both metrics, we observe statistically significant improvements in the fine-tuned version compared to the original (\textit{p-value} $< 0.01$).
Specifically, the median CC and MI values decreased by 9.09\% and 5.44\%, respectively, with effect sizes of $r = 0.54$ and $r = 0.56$.
These findings are consistent with our earlier results, despite focusing on different dimensions of quality.
However, it is important to note that ReCatcher evaluates a broader set of metrics, tailored to known issues in LLM-generated code~\cite{abbassi2025unveiling}.
The scripts used for this evaluation, including the results for each metric, are available in our online appendix~\cite{ReCatcherReplication}.
}

\rev{Overall, our findings suggest that while leveraging LLMs as judges presents a valid and promising approach for assessing the correctness and performance of code snippets, 
these models still exhibit limitations in terms of accuracy—limitations that ReCatcher can overcome.
Furthermore, in terms of cost, the single LLM-based evaluation previously reported cost approximately 0.94 USD.
However, when extrapolated to the full scope of our study—which includes two benchmarks and nine model versions—we estimate a total cost of approximately 48 USD.
}

\section{Discussion}
\label{sec:discussion}

\rev{In this section, we discuss some implications of our results.
First, we highlight some key topics that could benefit practitioners and researchers.
Second, we discuss further aspects of extending ReCatcher.}

\subsection{Key Takeaways}
This study highlights that while LLMs' code generation capability is continuously evolving, improvements and regressions occur across various code aspects. 
Trade-offs are common, where gains in one aspect may come at the expense of regressions in another. 
The most volatile code aspects we observed are \textit{General Logic}, \textit{Errors} (\textit{Syntax Error}, \textit{Missing Declaration/Import}) and  \textit{Execution Time}. 
Based on our findings, we recommend prioritizing these aspects in regression testing before migration. 
However, memory usage remained relatively stable across most model updates.
Comprehensive testing is essential before adopting a new model, as higher capability claims do not always translate to improvements in all code aspects. 
For instance, GPT-4o exhibited regressions compared to GPT-3.5-turbo, underscoring the need for empirical validation over assumed superiority.

Regarding the model evolution scenarios, for fine-tuning, selecting a training dataset that aligns with the target programming language is crucial. 
Mismatches between the fine-tuning dataset and the intended use case can introduce syntax errors and logic inconsistencies.   
For model merging, reviewing the training process of the merged models is essential, as the choice of merged models directly impacts code generation capability. 
To preserve code quality and minimize regressions, merging should prioritize models trained specifically for coding tasks, ideally within a domain relevant to the user's needs.

Regarding the applicability of \textit{ReCatcher}, we advocate that it provides a structured framework for practitioners and model maintainers to assess trade-offs introduced by model updates. 
By systematically analyzing regressions across multiple code aspects, ReCatcher can serve as a benchmarking tool or integrated into a leaderboard enabling continuous tracking of model performance over time, \rev{covering other aspects not limited to correctness.}

\subsection{Extending ReCatcher}
\rev{ReCatcher is developed with a modular, plugin-based architecture, enabling both practitioners and researchers to adapt and extend it for their specific needs. 
In this section, we provide details on how the \textit{Code Generator} and \textit{Test Suite} components can be extended to support advanced configurations and additional functionality.}


\rev{For the \textit{Code Generator} component, users can add new benchmarks by specifying the required information in the \texttt{constants} configuration file.
This includes the name of the benchmark and the path to its location (after download).
ReCatcher then loads the corresponding file and accesses the tasks (i.e., prompts) defined for code generation.
Currently, the framework supports benchmarks formatted in JSON.}
\rev{For the \textit{Test Suite} component, users can either modify existing metrics or introduce new ones aligned with the supported dimensions.
For instance, in the case of \textit{code duplication} checks, the default threshold is set to 10 tokens—i.e., two code fragments must share at least ten tokens to be considered duplicates.
This threshold can be modified in the configuration file.
To add new metrics (e.g., for static analysis issues), users need to implement a corresponding function (e.g., in \texttt{utils.py}) and ensure it is registered among the available static issue checks.}

\section{Related Work}
\label{sec: related_work}

Prior research on LLMs evaluation for code generation is limited in scope, overlooking code generation comprehensive comparisons and regression detection. ReCatcher fills this gap with a structured framework for LLMs code generation regression testing.

Although LLMs proved their capability in code generation~\cite{jiang2024survey}, their generated code often suffers from inefficiencies across multiple code aspects. 
LLMs for code generation are mainly evaluated using correctness scores (e.g., pass@k) or code structure similarity (e.g., CodeBleu)~\cite{ren2020codebleu}.
While these metrics assess syntactic and lexical similarities, they fail to capture essential software development concerns beyond correctness~\cite{zheng2024beyond}. 
Recent works have expanded LLM evaluation beyond correctness.
Siddiq et al.~\cite{10.1145/3643991.3645071} assessed GPT-generated code quality by analyzing the presence of code smells using static analysis tools.
However, their study focuses solely on static code properties, neglecting performance and lacking a structured evaluation framework for broader adoption. 
Similarly, Zheng et al.~\cite{zheng2024beyond} proposed RACE, which evaluates LLM-generated code across four dimensions: correctness, readability, maintainability, and performance. While RACE aligns with our objective, it lacks structured test formulations, making its coverage of inefficiency subcategories incomplete.
\rev{Furthermore, RACE primarily targets general aspects of code quality, relying on conventional
standard metrics, rather than challenges specific to LLM-generated code \cite{abbassi2025unveiling}.}
Despite these efforts, LLMs code generation capabilities comparisons still primarily rely on correctness metrics.
For instance, Tsai et al.~\cite{tsai2024codelessalignmore} compared original and fine-tuned models based only on correctness scores, while Liu et al.~\cite{liu2025sensmergingsensitivityguidedparameterbalancing} followed a similar approach after merging models.
Unlike previous works, \textit{ReCatcher} introduces a comprehensive regression testing framework that systematically compares LLM-generated code across identified inefficiencies in LLM-generated code~\cite{abbassi2025unveiling}.
More capable LLMs do not always guarantee improvements; they can introduce unexpected regressions~\cite{ma2024my}, highlighting the need for systematic regression testing in LLMs.

In traditional SE, regression testing ensures that changes—such as feature updates or bug fixes—do not introduce unintended behavior. This is typically achieved by re-executing unit tests~\cite{biswas2011regression}.
However, unlike traditional SE, LLMs are non-deterministic, making it difficult to define and apply unit tests consistently~\cite{ouyang2025empirical}.
To address regression risks in LLMs, RETAIN~\cite{dixit2024retain} was proposed as an interactive tool for guiding LLM regression testing. RETAIN includes two key components: (i) an interactive interface tailored for regression testing and (ii) an error discovery module that provides textual descriptions of output differences between LLMs and suggests prompt modifications to reduce inconsistencies.
While RETAIN is effective for general-purpose LLM regression testing, it focuses solely on textual differences, making it unsuitable for evaluating LLM-generated code. Code generation requires assessment beyond text comparison, considering structural, semantic, and executional properties that influence both functional correctness and non-functional aspects~\cite{naik2024limitations}.
\textit{ReCatcher} addresses this limitation by introducing a code-specific regression testing framework. It enables structured comparison of LLM-generated code across multiple quality dimensions, systematically detecting regressions that go beyond textual differences.

\section{Threats to Validity}
\label{sec: threats_validity}
\textit{Construct Validity.} The proposed test for detecting static code inefficiencies may introduce faults, potentially affecting the results. To address this, we relied on proven and widely used tools: Pylint~\cite{pylint} for static analysis and CPD~\cite{pmd} for duplication detection, which has been validated in previous research~\cite{gulabovska2019survey, groce2021evaluating, zakeri2023systematic, svajlenko2015evaluating}. 
For performance testing, we acknowledge that executing unit tests alone does not fully capture performance bottlenecks. To address this, \textit{ReCatcher} also supports testing using large inputs. 
For logical correctness, we recognize that unit tests may have limited coverage. However, this can be improved by generating additional tests using tools such as EvoSuite~\cite{fraser2011evosuite}.
\rev{Regarding our comparison with baseline solutions, while we selected baselines based on prior work and relevance, we acknowledge that stronger or additional baselines could further strengthen the evaluation. 
Moreover, relying on LLMs as automated judges introduces inherent risks, as these models may not always accurately assess the targeted properties. 
Future work will consider a broader set of comparative models.}

\textit{Internal Validity.} Non-determinism in LLM-generated outputs may introduce bias in our results. To mitigate this, we repeated the generation process 10 times. 
Additionally, we set the temperature to $0.1$ to limit randomness and reduce variability in outputs. For performance evaluation, each execution was repeated 5 times to minimize fluctuations. To ensure that performance measurements were not affected by external system interference, we conducted the assessments in an isolated environment with no other running processes. Instead of relying solely on average comparisons, we used \textit{Mann-Whitney U test} to ensure the reliability of our findings. This approach provides a more robust assessment of regression between models, accounting for variations beyond simple mean comparisons.
Another potential threat is the interdependence of test results. Certain inefficiencies, such as syntax errors and missing declarations/imports, directly impact logical correctness. This interdependency could misestimate logical correctness issues, as some logically correct code may not reach execution. 
Automatically assessing logical correctness in the presence of syntax errors and missing declarations is inherently challenging. However, if a code snippet requires manual intervention for basic fixes before execution, it is not fully reliable, making its logical correctness evaluation less relevant in practical scenarios.
\rev{Finally, when using LLMs as a judge, these models are inherently non-deterministic and may produce overconfident or inaccurate explanations, particularly for performance judgments. To mitigate this, we repeated evaluations multiple times to mitigate possible errors.}

\textit{External Validity.} Our study focuses on a specific set of models, analyzing their behavior under fine-tuning, merging, and new model releases. While these insights are valuable for understanding model regression, our findings may not generalize to LLMs with different architectures, training strategies, or fine-tuning datasets.
To evaluate fine-tuning and merging, we selected one fine-tuned and one merged variant per model. Selection was based on the highest number of downloads and well-documented model cards available on Hugging Face. While this ensures that the chosen models are widely used and sufficiently documented, other fine-tuned or merged variants may exhibit different behaviors.
Our findings are also influenced by the chosen benchmarks, which may limit generalizability. To mitigate this, we used two complementary benchmarks: HumanEval+~\cite{liu2024your}, a widely adopted benchmark for assessing LLM code generation performance, and BigCodeBench~\cite{zhuo2024bigcodebench}, which includes a diverse set of real-world programming tasks beyond algorithmic problem-solving. 
To enhance reproducibility and support research, we provide a replication package \cite{ReCatcherReplication} containing the generated code snippets, test results, and the \textit{ReCatcher} source code.

\textit{Conclusion Validity.} Our conclusions are based on quantitative analysis using automated tools, relying on Hugging Face model cards for information about model fine-tuning and merging. However, since not all model cards provide complete details, this could influence the interpretation of certain results. To mitigate this, we selected only highly downloaded Hugging Face models with comprehensive model cards that included details on fine-tuning datasets, merged models, and applied techniques. For OpenAI models, we conducted a thorough review of their website and research papers to gather the necessary information.

\section{Conclusion}
\label{sec: conclusion}
In this paper, we introduced \textit{ReCatcher}, the first systematic LLMs regression testing framework for python code generation. 
\textit{ReCatcher} enables a comprehensive regression testing between LLMs by assessing logical correctness, static code quality, and performance efficiency.
We used \textit{ReCatcher} to test regression across three scenarios: fine-tuning, merging, and model release on CodeLlama, DeepSeek-Coder, and GPT-family models. 
Our findings highlight key trends: (i) the dataset used for fine-tuning can introduce syntax regression, (ii) model merging with a foundation LLM may amplify overall regression, and (iii) GPT-4o exhibits regressions in handling missing imports compared to GPT-3.5-turbo, while GPT-4o-mini shows performance regressions in execution time.
\textit{ReCatcher} is open-sourced to promote further research and real-world adoption, assisting users in making informed decisions about model updates.  
\rev{Finally, when comparing \textit{ReCatcher} with baseline solutions, it shows that \textit{ReCatcher} presents higher accuracy regarding \textit{logical correctness} and \textit{performance}, while covering \textit{static code issues} particular to LLMs inefficiencies.}
As future work, we plan to extend \textit{ReCatcher} by integrating distribution-aware analysis for a more granular assessment of regressions. Additionally, we aim to refine regression analysis by considering task difficulty levels, providing deeper insights into LLM code generation capability.

\section*{Acknowledgements}
\label {acknowledgements}
We thank the anonymous reviewers for their valuable comments on improving an earlier version of this paper. 
This work is funded by the following organizations and companies: Fonds de Recherche du Quebec (FRQ), Natural Sciences and Engineering Research Council of Canada (NSERC), the Canadian Institute for Advanced Research (CIFAR), and the Canada Research Chairs Program. However, the findings and opinions expressed in this paper are those of the authors and do not necessarily represent or reflect those organizations/companies.

\section*{Data Availability}
\label{sec:data_availability}
To promote open science and facilitate reproducibility, we make all our artifacts available to the community.
This includes the proposed framework, ReCatcher, the scripts used for evaluation, and the LLM-generated code snippets during our study \cite{ReCatcherReplication, datasetZenodo}.
\bibliographystyle{ACM-Reference-Format}

\end{document}